\documentclass[twocolumn,showpacs,prb,amsfonts,amsmath,amssymb,floatfix]{revtex4} 
\usepackage{color}
\usepackage{hhline}
\usepackage{mathrsfs}
\usepackage{graphicx}
\usepackage{dcolumn}
\usepackage{bm}
\usepackage{multirow}
\usepackage{booktabs}
\usepackage{afterpage}
\usepackage{amsmath}

\begin{document}

\title{Robust flat bands in $R$Co$_5$ ($R=$ rare earth) compounds}

\author{Masayuki Ochi$^{1,2}$}\email{masayuki.ochi@riken.jp}
\author{Ryotaro Arita$^{1,2}$}
\author{Munehisa Matsumoto$^{3}$}
\author{Hiori Kino$^{3,4}$}
\author{Takashi Miyake$^{3,5}$}
\affiliation{$^1$RIKEN Center for Emergent Matter Science (CEMS), Wako, Saitama 351-0198, Japan}
\affiliation{$^2$JST ERATO Isobe Degenerate $\pi$-Integration Project, Advanced Institute for Materials Research (AIMR), Tohoku University, Sendai, Miyagi 980-8577, Japan}
\affiliation{$^3$ESICMM, National Institute for Materials Science, Tsukuba, Ibaraki 305-0047, Japan}
\affiliation{$^4$MANA, National Institute for Materials Science, Tsukuba, Ibaraki 305-0044, Japan}
\affiliation{$^5$Nanosystem Research Institute, RICS, AIST, Tsukuba, Ibaraki 305-8568, Japan}

\date{\today}
\begin{abstract}
The mechanism to realize the peculiar flat bands generally existing in $R$Co$_5$ ($R=$ rare earth) compounds is clarified by analyzing the first-principles band structures and the tight-binding model.
These flat bands are constructed from the localized eigenstates, the existence of which is guaranteed by the destructive interference of the intersite hopping among the Co-$3d$ states at the Kagom{\' e} sites and those between the Kagom{\' e} and honeycomb sites.
Their relative positions to other bands can be controlled by varying the lattice parameters keeping their dispersion almost flat, which suggests the possibility of flat-band engineering.
\end{abstract}

\maketitle

\section{Introduction}

The nature of flat bands has been attracting much attention from a variety of view points.
Ferromagnetism originating from the flat band has been intensively studied in some model systems~\cite{Lieb,Mielke,Mielke2,Tasaki,MielkeTasaki,Tasaki_review}.
The nearly flat band with a non-zero Chern number offers a unique playground of the fractional quantum Hall effect~\cite{QHE1,QHE2,QHE3,QHE4}.
An extremely large effective mass for the flat band affects the transport properties of solids and results in various unconventional phenomena such as the inverse Anderson transition~\cite{flat_Anderson}.
A sharp peak of the density of states (DOS) owing to partially flat dispersion is preferable for the thermoelectric devices to enhance their thermopower~\cite{thermo2,thermo3,pudding,AritaSeebeck}.

Owing to these various intriguing aspects, seeking flat bands in real materials is of significant importance for materials design.
One example of experimental realization of a flat band was reported for Cu(1,3-bcc)~\cite{Cu_Kagome,flat_material}, which can be described by a single-orbital model on the Kagom{\'e} lattice.
Another example is the tetragonal cuprate La$_4$Ba$_2$Cu$_2$O$_{10}$~\cite{La422_Hirabayashi,La422_LDA,Tasaki_review}, in which electronic states for a flat band play a central role in determining its magnetic order~\cite{La422_WK}.
Also other materials have been reported to possess flat bands~\cite{Shima1993,Wire,Arita_goinkan,goinkan2,IPOF}.

\begin{figure}
 \begin{center}
  \includegraphics[scale=.055]{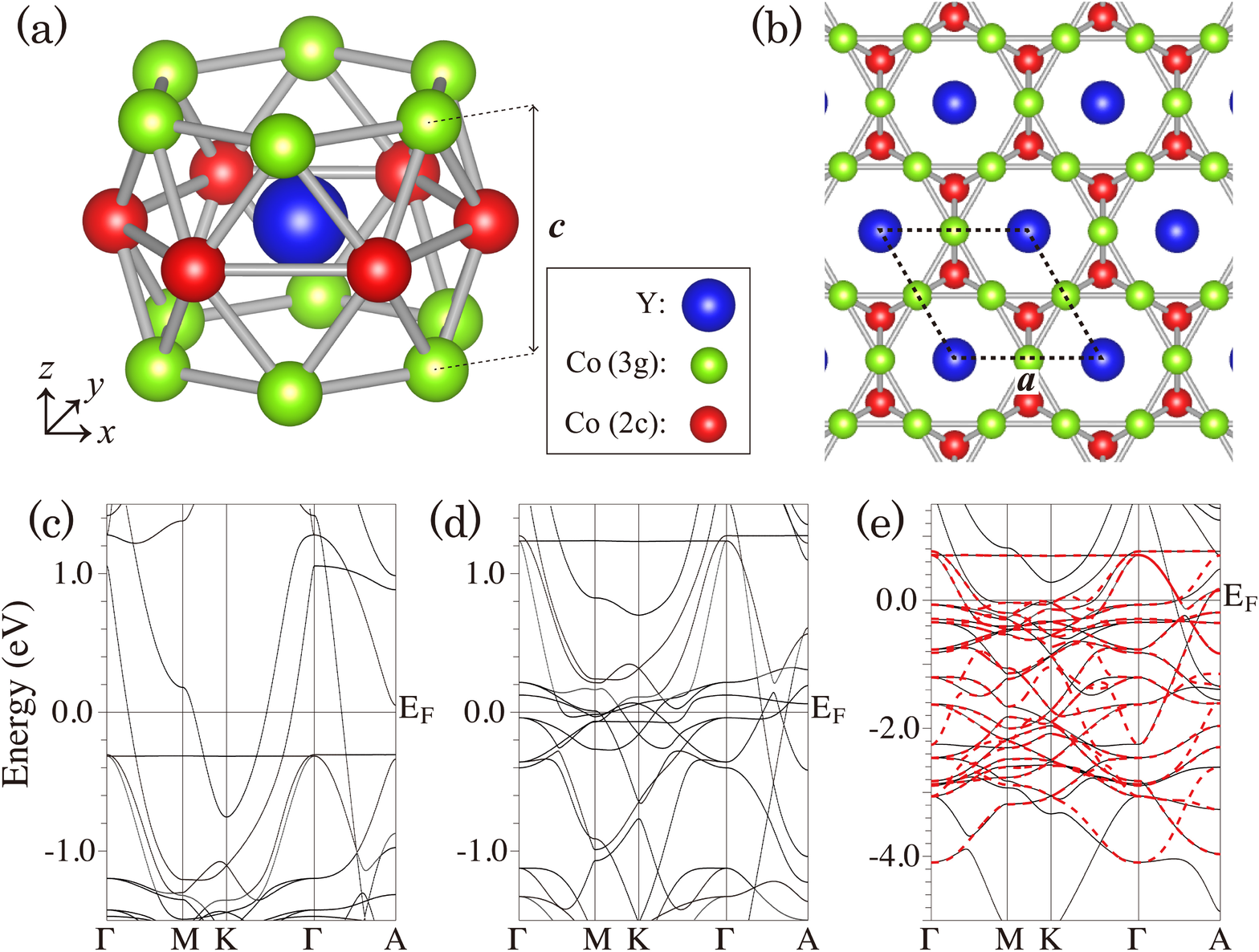}
  \caption{(Color online) (a) Crystal structure of YCo$_5$ and (b) its top view drawn using the VESTA software~\cite{VESTA}. Its first-principles band structures for the (c) majority and (d) minority spins. (e) Spin-unpolarized band structures obtained from the first-principles (solid black line) and the tight-binding-model (broken red line) calculations where the Fermi level of the latter is adjusted to have the flat band along the $\Gamma$--A line placed in the same position as that in the former.}
  \label{fig:1}
 \end{center}
\end{figure}

Recently, the intermetallic ferromagnet YCo$_5$ [Figs.~\ref{fig:1}(a) and \ref{fig:1}(b)], which is known for its large magnetic anisotropy, was pointed out to have peculiar flat bands~\cite{Nature_YCo5,PRB_YCo5}.
It was found that the system experiences the first-order Lifshitz transition~\cite{Lifshitz} and exhibits an unusual isomorphic lattice collapse with a sudden change of the magnetic moment when the Fermi level crosses the flat band [along the $\Gamma$--M--K--$\Gamma$ line in Fig.~\ref{fig:1}(c)] by applying pressure.
This flat band is observed in the whole $k_z=0$ plane and consists of Co-$3d_{xz}$ and $3d_{yz}$ states~\cite{PRB_YCo5}.
Similar flat bands were found in other $R$Co$_5$ ($R$=rare earth) compounds such as LaCo$_5$~\cite{PRB_YCo5}, SmCo$_5$~\cite{SmCo5_flat}, and also CePt$_5$~\cite{CePt5}, which has the same CaCu$_5$-type structure.
However, the mechanism to realize such ubiquitous flat bands has not been clarified yet.

In this paper, we reveal the origin of the flat bands in $R$Co$_5$ compounds and attribute their existence to the localized eigenstates realized by the destructive interference~\cite{Mielke,Mielke2,Tasaki,MielkeTasaki,Tasaki_review} of the inter-site hopping among Co-3$d$ orbitals on the Kagom{\'e}-honeycomb stacked structure.
These flat bands are affected very little by varying the lattice parameters while their relative positions to the other bands are altered.
Such robustness and controllability of the flat bands are crucial for the system to exhibit the first-order Lifshitz transition where the flat dispersion should be retained under pressure.
Our mechanism can be applied also to other CaCu$_5$-type structures with $d$ orbitals on Cu sites.
In this sense, this study demonstrates a general mechanism behind the flat dispersion for a wide range of materials.

\section{Computational details}

First-principles band structure calculations in this paper were performed using the \textsc{wien2k} code~\cite{wien2k}.
We used the Perdew-Wang local-spin-density approximation~\cite{PW92} and the full-potential linearized augmented plane-wave method. For simplicity, we do not include the spin-orbit coupling throughout this paper because it affects very little the flat bands.
Experimental lattice parameters ($a=9.313$ a.u.\ and $c/a=0.806$) are taken from Ref.~[\onlinecite{Expt2}].
The muffin-tin radii for Co and Y atoms, $r_{\rm Co}$ and $r_{\rm Y}$, were set to 2.34 and 2.07 a.u., respectively.
The maximum modulus for the reciprocal lattice vectors $K_{\rm max}$ was chosen so that $r_{\rm Y} K_{\rm max}$ = 9.00.

\section{Analysis on a tight-binding model}
\subsection{Construction of the model}

As was pointed out in Ref.~[\onlinecite{PRB_YCo5}], the flat bands in YCo$_5$ exist (i) on the $k_x=k_y=0$ line and (ii) in the $k_z=0$ plane [see the flat bands along the $\Gamma$--A and the $\Gamma$--M--K--$\Gamma$ lines, respectively, in Fig.~\ref{fig:1}(c)--(e)]. 
To analyze the origin of these flat bands, we constructed a tight-binding model consisting of Co-$3d$ states with all possible hopping paths taken into account.
Twenty-five maximally localized Wannier functions~\cite{Wannier1,Wannier2} were constructed from the Kohn-Sham states within the energy window [$-4.5$, $+1.5$] eV on a 6$\times$6$\times$6 $k$ mesh, where the energy is measured from the Fermi level.
We neglected Co-$4s$ and Y states because they were found to have no weight on these flat bands in our first-principles band structure calculations.
To construct the tight-binding model, we employed spin-unpolarized calculations because the flat bands also appear there and our story does not depend on whether the spin is polarized or unpolarized.
In Fig.~\ref{fig:1}(e), we can see that our tight-binding model reproduces the first-principles band structure well.

\subsection{General discussion for the localized eigenstate}

The Bloch states $\psi_{\mathbf{k}}(\mathbf{r})$ ($\mathbf{k}$: the crystal wave vectors) for the flat band are constructed as $\psi_{\mathbf{k}}(\mathbf{r})=\sum_{\mathbf{R}} e^{i\mathbf{k}\cdot\mathbf{R}}\phi(\mathbf{r}-\mathbf{R})$ where $\mathbf{R}$ runs all the lattice vectors and $\phi(\mathbf{r})$ is the localized eigenstate~\cite{Mielke,Mielke2,Tasaki,MielkeTasaki,Tasaki_review}.
To treat the flat bands in the subspace of the Brillouin zone, we instead consider $\sum_{\mathbf{R}\in S} e^{i\mathbf{k}\cdot\mathbf{R}}\varphi(\mathbf{r}-\mathbf{R})$ for a subspace $S$ and $\varphi(\mathbf{r})$ localized only for the directions included in $S$.
This summation yields the Bloch states only when $\varphi(\mathbf{r})$ periodically extends in $S^{\perp}$ with the same periodicity as the crystal.
For example, for the $k_x=k_y=0$ flat band, $S= \{ (0, 0, R_z)|R_z\in \mathbb{R}\}$ and the Bloch states are obtained as $\sum_{R_z} e^{ik_zR_z}\varphi(\mathbf{r}-\mathbf{R})$ where $\varphi(\mathbf{r})$ is localized only for the $z$ direction and periodically extends to the $x$ and $y$ directions.
We shall focus on showing the existence of such localized eigenstate, which guarantees the flat band dispersion.
Localized eigenstates were obtained by Fourier transform (in the restricted subspace) of the Bloch states in our tight-binding model.

\subsection{Localized eigenstate for the $k_x=k_y=0$ flat band}

\begin{figure}
 \begin{center}
  \includegraphics[scale=.07]{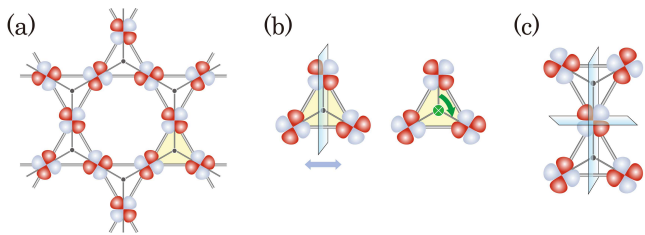}
  \caption{(Color online) (a) Schematic picture of the localized eigenstate for the $k_x=k_y=0$ flat band. (b) Reflection and $C_3$ rotation, which generate the $C_{3\mathrm{v}}$ group. (c) Two reflections on each Kagom{\'e} site.}
  \label{fig:flat1}
 \end{center}
\end{figure}

Figure~\ref{fig:flat1}(a) presents a schematic picture of the localized eigenstate $\varphi(\mathbf{r})$ for the $k_x=k_y=0$ flat band.
This state consists of Co-$3d_{xy}$ and $3d_{x^2-y^2}$ states, periodically extends to the $x$ and $y$ directions, and is confined in one Kagom{\'e} plane.
This state is the exact localized eigenstate of our tight-binding Hamiltonian $\hat{H}$ when neglecting second or higher-order nearest neighbor hoppings~\cite{interlayer} for the following reason: $\varphi(\mathbf{r})$ belongs to the irreducible representation A$_2$ of the symmetry group $C_{3\mathrm{v}}$, which is generated by the reflection and rotation shown in Fig.~\ref{fig:flat1}(b) [i.e. $\varphi(\mathbf{r})$ is antisymmetric for reflection and symmetric for rotation]. No $s$, $p$, or $d$ orbital on the honeycomb sites located on the rotational axis belongs to the same representation.
Thus the hopping integrals $\langle \varphi_{\mathrm{atom}}|\hat{H}|\varphi \rangle$ for all the atomic orbitals $\varphi_{\mathrm{atom}}(\mathbf{r})$ on the honeycomb sites vanish.
In addition, the atomic orbitals not included in $\varphi(\mathbf{r})$ (e.g. $d_{xz}$ and $d_{yz}$) on the Kagom{\'e} plane are also decoupled from $\varphi(\mathbf{r})$ by considering the antisymmetry of $\varphi(\mathbf{r})$ with respect to two reflections on each site shown in Fig.~\ref{fig:flat1}(c).
Along with such decoupling of the wave function, it is also necessary to prove that $\varphi(\mathbf{r})$ is an eigenstate of the Hamiltonian.
Note that any succession of $C_3$ rotations shown in Fig.~\ref{fig:flat1}(b) with different axes does not change $\varphi(\mathbf{r})$ and $\hat{H}$ at all, which means $\hat{H}\varphi(\mathbf{r})$ also satisfies this invariance.
Because arbitrary two sites of the Kagom{\'e} plane relate by such transformation, all the sites should have the same orbital weight in $\hat{H}\varphi(\mathbf{r})$.
Thus $\hat{H}\varphi(\mathbf{r})$ should be proportional to $\varphi(\mathbf{r})$.
Here all symmetries that $\varphi(\mathbf{r})$ satisfies should be also satisfied in $\hat{H}\varphi(\mathbf{r})$, which guarantees the decoupling between $\hat{H}\varphi(\mathbf{r})$ and the atomic orbitals not included in $\varphi(\mathbf{r})$.
Because the above discussion is solely based on symmetry, this flat band is ubiquitous in materials with the same structure having $d$ orbitals on the Kagom{\'e} sites~\cite{LaRh3B2,JPSJ_YCr6Ge6}. The same story also holds for the system where $p$ orbitals exist on the honeycomb sites such as YCr$_6$Ge$_6$~\cite{JPSJ_YCr6Ge6}.

\subsection{Localized eigenstate for the $k_z=0$ flat band}

\begin{figure}
 \begin{center}
  \includegraphics[scale=.095]{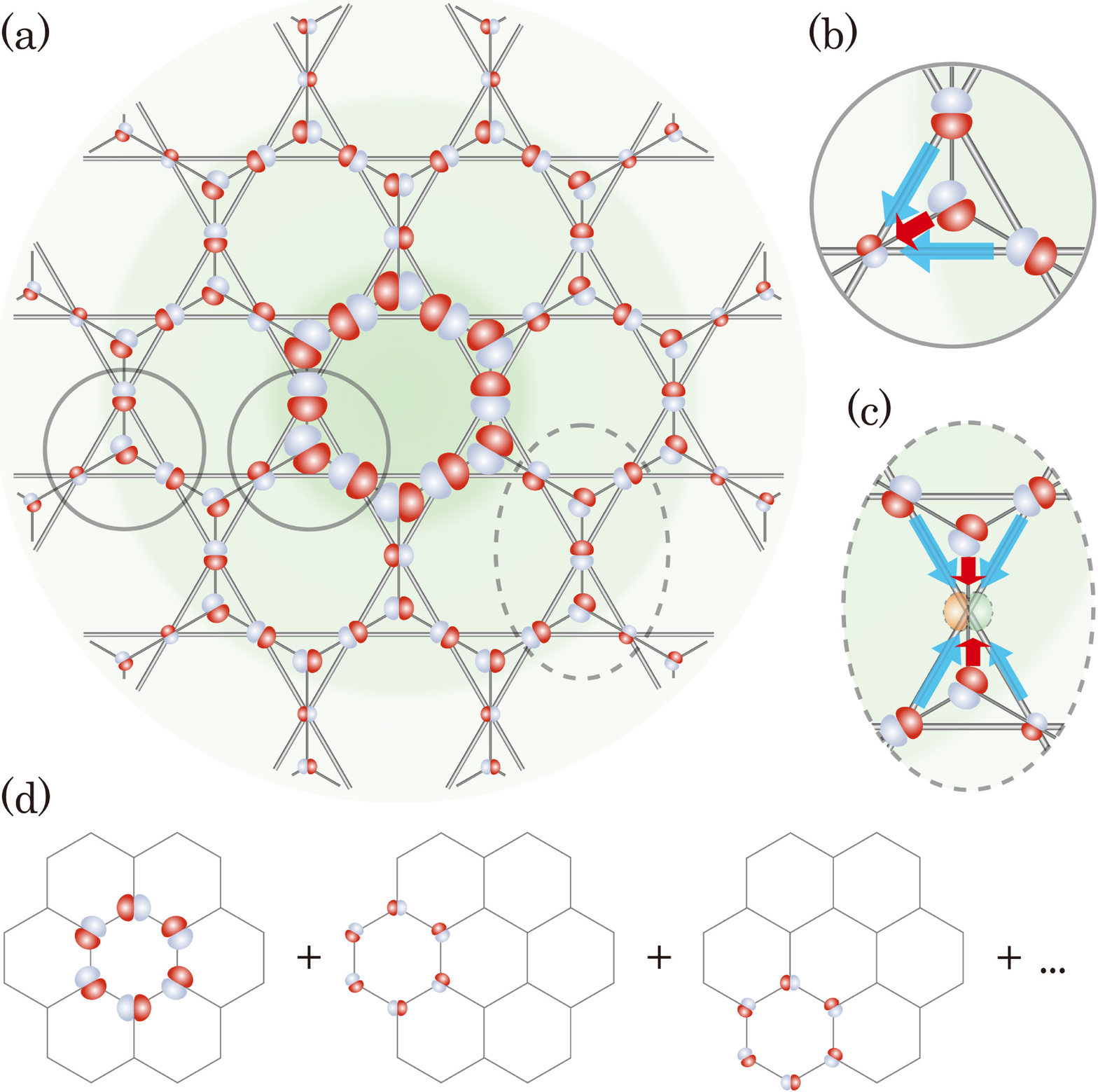}
  \caption{(Color online) (a) Schematic picture of the localized eigenstate for the $k_z=0$ flat band. The signs of all the orbitals at the middle between the Kagom{\'e} and the honeycomb planes are shown. The sizes of the orbitals are depicted just schematically and the same sizes only mean similar weights. Shaded background is a guide to the eye. Panels (b) and (c) show two characteristic interference processes to realize the localization where a different color of an arrow means a different sign of the hopping amplitude. (d) Linear combination of the localized states in the honeycomb lattice~\cite{honeycomb_flat}.}
  \label{fig:flat2}
 \end{center}
\end{figure}

Figure~\ref{fig:flat2}(a) presents a schematic picture of the localized eigenstate with a damping tail~\cite{surface_flat} for the $k_z=0$ flat band.
This state consists of Co-$3d_{xz}$ and $3d_{yz}$ states~\cite{PRB_YCo5} both on Kagom{\'e} and honeycomb planes, and periodically extends to the $z$ direction with period $c$ defined in Fig.~\ref{fig:1}(a).
Atomic orbitals except $3d_{xz}$ and $3d_{yz}$ states are not involved owing to the mirror symmetry with respect to the $xy$-plane where each orbital is placed.
The destructive interference shown in Fig.~\ref{fig:flat2}(b) is the origin of an interesting step-by-step damping of the orbital weights in Fig.~\ref{fig:flat2}(a).
The relative weights of the large, middle, and small orbitals in Fig.~\ref{fig:flat2}(a) are about 31 : 4 : 1 in our calculation.
The fact that cancellation is not required to be perfect makes the localization insensitive to the values of the tight-binding parameters.
We observed that, as shown in Fig.~\ref{fig:flat2}(c), the net hopping to the atomic orbital depicted in the center of this figure vanishes, and then this orbital is not involved with the localized eigenstate.
Thanks to this fact and symmetry of the localized eigenstate $\varphi(\mathbf{r})$, the atomic orbitals $\varphi_{\mathrm{atom}}(\mathbf{r})$ except those depicted in Fig.~\ref{fig:flat2}(a) satisfy $\langle \varphi_{\mathrm{atom}}|\hat{H}|\varphi \rangle=0$.

\begin{figure}
 \begin{center}
  \includegraphics[scale=.071]{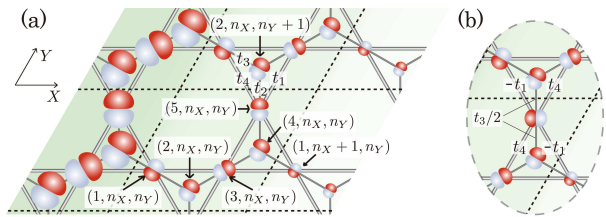}
  \caption{(Color online) (a) Exponentially decaying eigenstate with the definition of the site indices and the hopping amplitudes $t_l$ ($l=1,2,3,4$). (b) Hopping amplitudes for Fig.~\ref{fig:flat2}(c).}
  \label{fig:4}
 \end{center}
\end{figure}

Owing to the complexity of Fig.~\ref{fig:flat2}(a), it is difficult to prove exactly the existence of the localized eigenstate. Thus we shall instead show the existence of the exponentially decaying eigenstate in a simplified situation as depicted in Fig.~\ref{fig:4}(a), which is expected to mimic $\varphi(\mathbf{r})$ far from its center. To begin with, we define some variables.
Each site on the lattice is specified as $i=(m_i,n_{Xi},n_{Yi})$ with $n_{Xi}$ and $n_{Yi}$ denoting the unit cells and $m_i\ (=1,2,3,4,5)$ the sites inside the cell as shown in Fig.~\ref{fig:4}(a).
Since we treat the $k_z=0$ states, each atomic orbital here is defined as an infinite sum over the $z$ direction with the same phase, and our problem is converted to that on the two-dimensional lattice.
In other words, we move on to the subspace spanned by the periodic functions to the $z$ direction with period $c$.
We consider only $d_{xz}$ and $d_{yz}$ states thanks to the mirror symmetry described before, and so each site has two orbital degrees of freedom.
We define an orbital index $P=1$ for the orbitals depicted in the $(n_X,n_Y)$ cell of Fig.~\ref{fig:4}(a) and $P=2$ for those rotated 90 degrees counterclockwise in the $xy$-plane.
For example, the creation operator for the orbital depicted in the center of Fig.~\ref{fig:4}(b) is $\hat{c}_{(5,n_X,n_Y)}^{2\dag}$.
For simplicity, we omit the spin index. The Hamiltonian is
\begin{equation}
\hat{H} = \sum_{i, P} \epsilon_i^P \hat{c}_i^{P\dag} \hat{c}_j^P + \sum_{\langle i j \rangle, P, Q} \left( t_{ij}^{PQ}\hat{c}_{i}^{P\dag}\hat{c}_j^Q + H.c. \right),
\end{equation}
where $i$ and $j$ denote the site indices, $P$ and $Q$ the orbital indices, and $\langle ... \rangle$ means a nearest-neighbor Co(3g)-Co(3g) or Co(3g)-Co(2c) pair.
Here we neglect other distant hopping processes.
Using the symmetry of the crystal, we can express all the hopping amplitudes $t$ with only four parameters $t_l$ ($l=1,2,3,4$) defined in Fig.~\ref{fig:4}(a).
Some examples are shown in Fig.~\ref{fig:4}(b). The weight of the atomic orbital $P$ on the site $(m_i,n_{Xi},n_{Yi})$ for the eigenstate of our Hamiltonian is denoted as $x_{(m_i,n_{Xi},n_{Yi})}^P$.

We shall prove the existence of the eigenstate that satisfies, for all $(m_i,n_{Xi},n_{Yi})$, (i) $x_{(m_i,n_{Xi},n_{Yi})}^P=\alpha^{n_{Xi}} x^P_{(m_i,0,0)}$ $(P=1, 2)$ for some complex value $\alpha$ and (ii) $x_{(m_i,n_{Xi},n_{Yi})}^2=0$.
Such exponentially decaying eigenstate as depicted in Fig.~\ref{fig:4}(a) is considered to guarantee a decay of the localized eigenstate as presented in Fig.~\ref{fig:flat2}(a) in the region far from the center of the localized eigenstate in an approximate sense.
As long as $|\alpha|\neq 1$, we can obtain a decaying eigenstate to the right or left direction.

Because the Hamiltonian and the translational operators to the $X$ and $Y$ directions with the same range as the unit cell commute, we can immediately obtain the simultaneous eigenstate of these three operators, which satisfies (i) $x_{(m_i,n_{Xi},n_{Yi})}^P=\alpha^{n_{Xi}} x^P_{(m_i,0,0)}$ where the eigenvalues of the translational operator to the $X$ and $Y$ directions are $\alpha$ and $1$, respectively.
This is the Bloch's theorem for the case where the periodic boundary condition is not imposed.
Hereafter, we denote $x_{m_i}\equiv x_{(m_i,n_{Xi},n_{Yi})}^1$ for simplicity.

The following two conditions suffice to realize (ii) $x_{(m_i,n_{Xi},n_{Yi})}^2=0$:
\begin{gather}
x_3=x_5,\label{eq:symmetry}\\
-(1+\alpha)t_1x_1+\frac{t_3}{2}(x_2+x_4)+2t_4x_3=0.\label{eq:alpha}
\end{gather}
because the hopping integral between the eigenstate and the $P=2$ orbital on each site becomes exactly zero by these conditions.
In particular, Eq.~(\ref{eq:alpha}) corresponds to the interference of the hopping processes as shown in Fig.~\ref{fig:flat2}(c) [see Fig.~\ref{fig:4}(b) for the hopping parameters].

The eigenvalue equation for the Hamiltonian is
\begin{align}
\epsilon_{k1} x_1 + t_3 x_2 + 2(1+\alpha^{-1})t_4 x_3 + \alpha^{-1}t_3 x_4 &= \lambda x_1, \label{eq:1eigen}\\
t_3 x_1 + \epsilon_h x_2 + 2t_2 x_3 &= \lambda x_2,\label{eq:2eigen}\\
2(1+\alpha)t_4 x_1 + 2t_2 x_2 + (\epsilon_{k3}+2t_1)x_3 + 2t_2 x_4 &= \lambda x_3,\label{eq:3eigen}\\
\alpha t_3 x_1 + 2t_2 x_3 + \epsilon_h x_4 &= \lambda x_4,\label{eq:4eigen}
\end{align}
where $\lambda$ is the eigenenergy and the onsite energies are denoted as $\epsilon_{h}\equiv \epsilon_{(2,n_X,n_Y)}^1=\epsilon_{(4,n_X,n_Y)}^1$, $\epsilon_{k1}\equiv \epsilon_{(1,n_X,n_Y)}^1$, and $\epsilon_{k2}\equiv \epsilon_{(3,n_X,n_Y)}^1=\epsilon_{(5,n_X,n_Y)}^1$.
The left-hand sides of these equations are the orbital weights on $m_i=1, 2, 3, 4$ sites of the eigenstate applied by the Hamiltonian.
Equations~(\ref{eq:alpha}), (\ref{eq:2eigen}), and (\ref{eq:3eigen}) yield
\begin{gather}
\alpha=-1+\frac{1}{t_1x_1}\left[\frac{t_3}{2}(x_2+x_4)+2t_4x_3\right],\label{eq:alpha2}\\
x_1=-\frac{\epsilon_h-\lambda}{t_3}x_2-\frac{2t_2}{t_3}x_3,\label{eq:x1}\\
x_2=-\frac{(\epsilon_{k3}+2t_1-\lambda)t_1+4t_4^2}{2t_1t_2+t_3t_4}x_3-x_4.\label{eq:x2}
\end{gather}
By substituting Eqs.~(\ref{eq:alpha2})--(\ref{eq:x2}) into Eq.~(\ref{eq:4eigen}), we obtain
\begin{align}
&2\left(2t_2 + \frac{t_3t_4}{t_1}\right)^2\nonumber \\
&- \left(\epsilon_h+\frac{t_3^2}{2t_1}-\lambda\right) \left( \epsilon_{k3}+2t_1+4\frac{t_4^2}{t_1}-\lambda \right)=0,\label{eq:lambda}
\end{align}
which determines $\lambda$ as a function of the tight-binding parameters.
By substituting this and Eqs.~(\ref{eq:alpha2})--(\ref{eq:x2}) into Eq.~(\ref{eq:1eigen}), we obtain a quadratic equation for $x_4$. Because it is a quadratic equation, it always has solution(s).
This proof is valid for {\it arbitrary} values of the tight-binding parameters.
Essentially, a degree of freedom for $\alpha$ allows the eigenstate to satisfy the condition, Eq.~(\ref{eq:alpha}).

Beyond our simplification, the nearest-neighbor Co(2c)-Co(2c) hopping except the $\delta$-bonding contribution, ($dd\delta$) in the Slater-Koster parametrization~\cite{SlaterKoster}, can be easily taken into consideration and does not affect our conclusion because it just shifts the onsite energy of the honeycomb sites for the following reason.
As shown in Fig.~\ref{fig:flat2}(d), orbital weights on the honeycomb planes for $\varphi(\mathbf{r})$ can be constructed as the linear combination of the localized eigenstates in the honeycomb lattice~\cite{honeycomb_flat}.
Thus applying a part of our Hamiltonian, that is, the onsite terms for the honeycomb sites and the inter-honeycomb hopping terms, to $\varphi(\mathbf{r})$ just leads to the multiplication by the eigenvalue of the localized eigenstate of the honeycomb lattice.
This means that one can take these hopping terms into account only by replacing the onsite energy of the honeycomb sites with the eigenvalue of the localized eigenstate of the honeycomb lattice. 
Neglected ($dd\delta$) and other distant hopping processes have minor effects on our analysis because they have smaller amplitudes than the hopping processes taken into consideration here.
Actually we observed that the localized state shows some tilting of the outer small-weighted orbitals compared with our schematic picture in Fig.~\ref{fig:flat2}(a), which can be ascribed to such hopping processes and a simplified shape of the analyzed damping state in Fig.~\ref{fig:4}(a) from the real concentric damping of the localized eigenstate in Fig.~\ref{fig:flat2}(a).

\section{First-principles band structures with different lattice parameters}

\begin{figure}
 \begin{center}
  \includegraphics[scale=.065]{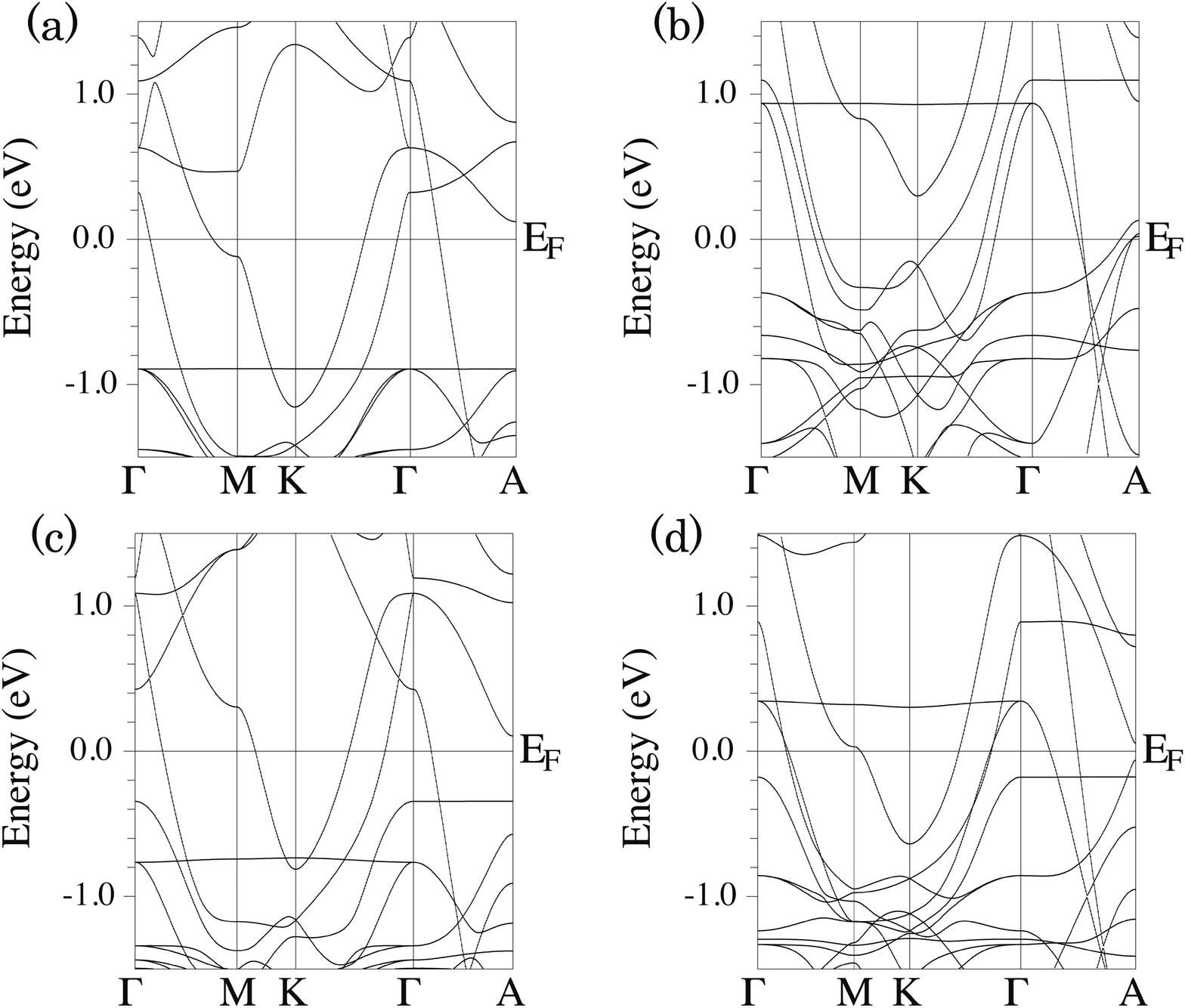}
  \caption{The YCo$_5$ band structures of the majority spin with lattice parameters (a) increased or (b) decreased by 10\% isotropically, and those with only the $c$ axis (c) increased or (d) decreased by 10\%.}
  \label{fig:5}
 \end{center}
\end{figure}

Figure~\ref{fig:5} presents the band structures of YCo$_5$ with lattice parameters (a) increased  or (b) decreased by 10\% isotropically, and those with only the $c$ axis (c) increased or (d) decreased by 10\%.
Although similar investigation using a small variation such as about 1\% for $c/a$ was performed in previous studies~\cite{Nature_YCo5,PRB_YCo5}, it is surprising that the flat dispersion is almost completely retained by such large variation of the lattice parameters.
The relative positions of the flat bands to the other bands and the Fermi level are different among these figures and thus are controllable by varying the lattice parameters.
Robustness presented here is consistent with the fact that other $R$Co$_5$ compounds~\cite{PRB_YCo5,SmCo5_flat} and also CePt$_5$~\cite{CePt5}, which has the same CaCu$_5$-type structure, were reported to have the same flat bands.

\section{Summary}

In summary, we found that the realization of the flat bands in $R$Co$_5$ compounds can be attributed to the existence of the localized eigenstate defined in particular directions.
Because the localized eigenstate for the $k_z=0$ flat band has an overlap with the neighboring localized eigenstate (i.e., that translated with respect to the $x$ and $y$ directions), the existence of the $k_z=0$ flat band can contribute to the in-plane ferromagnetic correlation~\cite{Lieb,Mielke,Mielke2,Tasaki,MielkeTasaki,Tasaki_review}.
It is also interesting that the non-trivial destructive interference is observed on the structure of the Kagom{\'e}-honeycomb network that can be constructed as a line graph~\cite{Mielke,Mielke2} of the hexagonal prism.
Their flat dispersion is very robust against the variation of the lattice parameters whereas their relative position to the other bands and the Fermi level can be controlled by it, which offers the possibility of flat-band engineering.

\acknowledgements
We appreciate helpful discussions with Kiyoyuki Terakura and valuable comments from Yoshihiro Gohda.


\begin{thebibliography}{999}
\bibitem{interlayer} A small coupling among the localized states in different Kagom{\'e} planes introduces a slight dispersion for the $k_x=k_y=0$ flat band, but the calculated bandwidth is about 2.5 meV in our tight-binding model, which is very small.
\end{thebibliography}
\end{document}